\documentclass[prd,nofootinbib,twocolumn]{revtex4-2}
\usepackage{graphicx, epsfig}
\usepackage{color}
\usepackage{mathrsfs}
\usepackage{bm}
\usepackage{amsmath,amssymb,yhmath}
\usepackage[caption=false]{subfig}
\usepackage{hyperref}
\usepackage[normalem]{ulem}
\usepackage[scr=boondoxo]{mathalpha}
\usepackage{bigints}

\newcommand{\be}{\begin{equation}}
\newcommand{\ee}{\end{equation}}
\newcommand{\ba}{\begin{eqnarray}}
\newcommand{\ea}{\end{eqnarray}}
\newcommand{\half}{\frac{1}{2}}
\newcommand{\nn}{\nonumber}
\newcommand{\dw}{{{}_{\rm DW}}}

\begin{document}
\title{Domain wall annihilation -- a QFT perspective}
\author{Oriol Pujol\`as$^{1}$}
\author{George Zahariade$^{1}$}
\affiliation{$^{1}$Institut de F\'isica d'Altes Energies (IFAE)  and The Barcelona Institute of Science and Technology (BIST)\\ 
Campus UAB
08193 Bellaterra (Barcelona), Spain.
}

\begin{abstract}
\noindent

Domain wall networks in the early universe, formed upon spontaneous breaking of a discrete symmetry, have a rich impact on cosmology. Yet, they remain somewhat unexplored. 
We introduce a new analytic strategy to understand better the domain wall epoch, from formation to annihilation.
Our method includes a quantum field theoretical treatment of the initial state at domain wall formation, as well as of the time evolution. 
We find that the domain wall area density for a network with biased initial condition in $d+1$ dimensional flat spacetime evolves as $t^{-1/2}\,\exp\big(- (t/t_{ann})^{d/2}\big)$. 
We comment on the relation between this and previous results obtained in condensed matter and in cosmology. 
The extrapolation of this law to an expanding universe applies to networks that are close to the domain wall `gas' limit.

%

\end{abstract}

\maketitle


\section{Introduction}
\label{introduction}

Domain wall (DW) networks are interesting probes of physics beyond the Standard Model (SM). DWs form whenever a discrete symmetry is spontaneously broken. A variety of SM extensions feature DWs, among others, axion models. 
In the early universe, any phase transition with spontaneous breaking of a discrete symmetry results in a cosmological DW network, which has a strong impact on cosmology, see \cite{Vilenkin:2000jqa,Vachaspati:2006zz} for reviews.
DW networks typically attain a {\em scaling} or {\em self-similar} regime, where at any time $t$ the network is composed of $O(1)$ Hubble-sized walls, and so the average network density is $\rho_\dw\sim\sigma_\dw / t$, with $\sigma_\dw$  the DW tension.
Thus DWs dilute more slowly than matter, bringing about the so-called `DW problem'. 
In practice, the DW problem is avoided if i) $\sigma_\dw$ is tiny ($\sigma_\dw\lesssim MeV^3$, \cite{Zeldovich:1974uw,Lazanu:2015fua}), or ii) the network annihilates at some time, $t_{ann}$. 

The interest in DW networks is strengthened because it is possible nowadays to search for the gravitational wave (GW) signal that they generate in GW observatories. The signal amplitude scales like $\sigma_\dw^2$ so the loudest signals necessarily correspond to networks that annihilate at some point. Moreover, DWs also arise in (post-inflationary) axion models, where a hybrid DW-string network is formed below the QCD  epoch and its annihilation affects the final axion dark matter abundance.
This renders the annihilating networks of particular interest, and considerable work has been dedicated to them recently \cite{Hiramatsu:2010yz,Hiramatsu:2010yn,Avelino:2010qf,Kawasaki:2011vv,Hiramatsu:2013qaa,Sousa:2015cqa,Krajewski:2016vbr,Nakayama:2016gxi,Martins:2016lzc,Krajewski:2017czs,Saikawa:2017hiv,Gelmini:2020bqg,Takahashi:2020tqv,Babichev:2021uvl,Dunsky:2021tih,ZambujalFerreira:2021cte,Ramberg:2022irf,Gelmini:2022nim,Ferreira:2022zzo,Jiang:2022svq,Blasi:2022woz,Beyer:2022ywc,Gonzalez:2022mcx}. 




Various mechanisms are known to result in the annihilation of DWs. 
To illustrate the various options, let us consider the discrete symmetry to be $\mathbb{Z}_2$ acting on a real scalar field as $\phi\to-\phi$. For concreteness we can assume
\be\label{V}
V(\phi)=\frac{\lambda}{4}\phi^4 + \frac{m^2(t)}{2} \phi^2~.
\ee
The potential $V(\phi)$ is $\mathbb{Z}_2$ symmetric and at some time $m^2(t)$ becomes negative and $V(\phi)$ develops a double-well shape with degenerate minima. There are three basic types of annihilation mechanisms (arising naturally in different models):
\begin{itemize}
\item {\em symmetry restoration}: the potential returns to a single $\mathbb{Z}_2$ preserving vacuum and consequently the DWs disappear. This case does not require any further breaking of $\mathbb{Z}_2$. However, having a DW finite epoch would require that the mass term becomes first tachyonic and later on positive. We won't discuss this option in this work.
\item {\em pressure bias}: consisting in the addition of small  explicit $\mathbb{Z}_2$-breaking terms in the potential that uplift the vacuum degeneracy.  The simplest option corresponds to adding to the potential the lowest dimension $\mathbb{Z}_2$-breaking operator, $\phi$. The potential difference in the two minima, $\Delta V$, results in a pressure pushing the walls towards the false vacuum regions.
\item {\em population bias}: more than including explicit symmetry breaking terms in the Lagrangian, the  symmetry is broken by assuming an uneven initial distribution of the degenerate vacua. This can be realized for instance by having the field displaced from $\phi=0$ at the time of the DW forming transition. 
\end{itemize}

The main focus of this work is network annihilation for biased networks, paying particular attention to how the annihilation proceeds as a function of time. 
One expects that the annihilation translates into an exponential suppression of the DW average density compared to the scaling regime. In other words,   
\be\label{S}
\rho_\dw = \frac{\sigma_\dw}{t}  \, S(t)
\ee
with $S(t)$ an exponentially decaying function over some typical annihilation scale $t_{ann}$.


The annihilation of simple  $\mathbb{Z}_2$ DW networks has been discussed before \cite{Ohta:1982zz,Coulson:1995nv,Hindmarsh:1996xv,Larsson:1996sp,Correia:2014kqa,Correia:2018tty,Krajewski:2021jje}, both analytically and numerically (using discretized field theory simulations). 
Early numerical simulations of models with population bias in an expanding universe were performed in \cite{Coulson:1995nv}. They were fitted to a phenomenological form
\be\label{coul}
S_{\rm CLO}=\exp\left( - \eta/\eta_{ann} \right)
\ee 
with $\eta=\int dt/a(t)$ the conformal time. The simulation results fitted reasonably well \cite{Coulson:1995nv}.

Exploiting the methods developed previously within solid state physics \cite{Ohta:1982zz}, Hindmarsh obtained  quite a remarkable result: the DW network energy density in an expanding $d+1$ dimensional universe should be suppressed by \cite{Hindmarsh:1996xv}\footnote{The suppressions \eqref{coul} or \eqref{hind} are reminiscent of percolation theory \cite{Stauffer:1978kr,Lalak:1993bp,Coulson:1995nv,Vilenkin:2000jqa}, where the number of DWs of different sizes formed at the symmetry breaking transition was found to scale with nontrivial powers of the size in the exponential factor. Let us just emphasize that Eqs. \eqref{coul}, \eqref{hind} refer to the network evolution in time, sufficiently late after the transition.}
\be\label{hind}
S_{\rm H}= \exp\left[ - \; U^2 \left(\frac{\eta}{\eta_i}\right)^{\hspace{-1mm} d}\, \right]~.
\ee
Here $\eta_i$ is some initial time, when the network is still self-similar, and
%
$U$ is a dimensionless measure of the bias, as a deviation from $Z_2$ symmetry in the (Gaussian) distribution of the field.
Note that the quadratic dependence in $U$ can be anticipated, since annihilation must take place independently of its sign. 
Writing the exponent as $-\eta^d/\eta_{ann}^d$, one identifies that the annihilation time scales with bias as $\eta_{ann} = \eta_i\,U^{-2/d}$.

As it stands, Eq.~\eqref{hind} is intended to apply for population bias. (The method does not include any pressure term in the equation of motion for the walls.) However, the decay \eqref{hind} has been checked against numerical simulations and it turns out to work well for  pressure bias and not so much for population bias. 
Indeed, various numerical simulations of both pressure and population biases were performed in \cite{Larsson:1996sp} and \cite{Correia:2014kqa,Correia:2018tty},  confirming that pressure bias follows \eqref{hind} but population bias instead decays less rapidly.

In an attempt to better understand the annihilation process, in the present work we aim at solving for the network evolution and annihilation analytically `from scratch' in the quantum field theory (QFT). For the moment we will restrict to the $\mathbb{Z}_2$ model and to flat space. 

~\\[-4mm]

Let us open a parenthesis to emphasize that there is a physical process sensitive to the precise form of the exponential suppression $S(t)$. This is the formation of primordial black holes (PBHs) from the collapse of the DW network \cite{Ferrer:2018uiu}. (See also \cite{Garriga:2015fdk,Deng:2016vzb} for a similar PBH formation mechanism from DWs). In essence, the network annihilation can be pictured as the reorganization of the DW shapes so that they form closed structures similar to false vacuum pockets. Once a closed DW fits into a Hubble volume, it collapses under the effect of both the tension $\sigma_\dw$ and the vacuum energy difference $\Delta V$. 
If the closed DWs shrink to small enough size, they can form black holes (BHs). The criterion for formation is that the collapsing DW fits within the Schwarzschild radius associated to it. This can be estimated from the total mass stored by the wall initially, when it is Hubble-sized and roughly at rest, {\em i.e.}, $R_S \sim (\sigma_\dw t^2 +\Delta V \, t^3 )/M_P^2$. The closer this is to the Hubble scale, $t$, the more likely it will be to form PBHs.
BH formation then strongly depends on the `figure of merit' \cite{Ferrer:2018uiu} defined as the ratio of the Schwarzschild to Hubble scales, $p=R_S / t $. It's easy to see that the condition $p\gtrsim 1$ actually coincides with DW domination. This must be avoided
for a viable cosmology. The largest value $p$ reaches is at $t_{ann}$, so we must have $p_{ann}\ll1$. At $t_{ann} = \sigma_\dw / \Delta V$ both bulk and surface terms contribute equally to the total DW mass so one has, in radiation domination, 
$p_{ann} \sim \Delta V / T^4_{ann}$. 
Avoidance of DW domination, $p_{ann}\ll1$, is certainly feasible but seems an insurmountable obstacle to produce PBHs. However,  this is not the end of the story. The reason is that the DW network contains closed walls of various sizes. The bigger DWs in the network from the tail of the distribution (called `late birds' in \cite{Ferrer:2018uiu}), only fit into a Hubble volume later on, at temperatures  $T_* < T_{ann}$. The figure of merit for them becomes very quickly enhanced, $p_*  = p_{ann} (T_{ann} / T_*)^4$, possibly above the collapse threshold $p_*\gtrsim1$ (without running into DW domination because this only happens in a small fraction of the Hubble volumes). So, sufficiently late birds can form PBHs, and their fraction is encoded in the exponential suppression in $\rho_\dw$. The details in the exponent are, then, important. 

%
%

Returning to our computation, our method to solve for the network annihilation is simply the extension of \cite{Mukhopadhyay:2020xmy,Mukhopadhyay:2020gwc} to include the bias.  Ref. \cite{Mukhopadhyay:2020xmy,Mukhopadhyay:2020gwc} discuss the formation of  walls in $1+1$ dimensions at the $\mathbb{Z}_2$ symmetry breaking transition (when $m^2(t)<0$) and their subsequent behaviour. At the transition, a collection of kinks and antikinks are produced, which later on collide and annihilate. 
The result, in the absence of bias and in flat space, is that the kink-antikink network (or `plasma'), quickly  sets into a self-similar evolution, in which the kink/antikink number density scales as \cite{Mukhopadhyay:2020xmy,Mukhopadhyay:2020gwc}
\be\label{nK1+1}
n_{\rm K} \sim \frac{1}{\sqrt t}~.
\ee
Notice that this is a  diffusive behaviour: the typical kink-antikink separation $L\equiv1/n_{\rm K}$ scales like $t^{1/2}$, we return to this below.

This result is obtained from a full QFT treatment, which relies mostly on the following observation \cite{Mukhopadhyay:2020xmy,Mukhopadhyay:2020gwc}: across the $\mathbb{Z}_2$ breaking transition, the kinks' positions and their evolution are already well defined at the free theory, $\lambda=0$, level. Walls can be defined simply as places where $\phi$ changes sign. 
It is then possible to track the motion of walls by keeping only the quadratic terms in the Lagrangian throughout the $\mathbb{Z}_2$ breaking transition where the mass term changes sign.

To make clear that we will be working in this free field approximation, we introduce the terminology {\em precursor kinks} or {\em precursor domain walls} understood as the positions/surfaces of the zeros of the free fluctuating (tachyonic) field. Precursor walls are related to the standard DWs, but differ in important aspects too. In the discrete symmetry breaking transition, first precursor walls are formed. Then, as the field grows to nonlinear values they eventually become standard domain walls. Once formed, standard DWs obey the Nambu Goto (NG) equation to a good approximation. Precursor walls, instead, do not obey the NG equation.

Our results are relevant for cosmological models accommodating a long precursor wall network epoch. (A wall precursor regime can be realized physically, for instance, if the scalar mass scale near the symmetry breaking transition is much lighter than the Hubble rate $H$ at the symmetry breaking transition.) The full computation in the case of an expanding universe is beyond the scope of the present work, but we shall make some comments. 

Working at the level of precursor walls will serve as an illustrative exercise to understand the details of the network annihilation. 
As we will see, already in flat space precursor wall networks follow a self-similar scaling regime -- of diffusive type similar to \ref{nK1+1}. 
An advantage of this QFT formalism (that can be exported to the cosmological case) is that the initial state for the network is properly described as the quantum vacuum of the field.

%
%

In the present work we shall do a first step in this analysis by analyzing the statistics of annihilating precursor wall networks in flat space in presence of a population bias.
In other words, our goal is to extend the result of \cite{Mukhopadhyay:2020xmy,Mukhopadhyay:2020gwc} to a model with bias, and to $d+1$ dimensions.
We add the bias as a $\mathbb{Z}_2$ breaking term in the Lagrangian. The simplest (and most relevant) operator is linear in the field $\phi$ with a possibly time dependent coefficient. With no loss of generality, the resulting quadratic potential with linear bias is
\be\label{bias}
V(\phi)\simeq\half m^2(t) (\phi-\delta\phi(t))^2 
\ee
and the c-number $\delta\phi(t)$ encodes the bias directly as the position of the maximum (for $m^2<0$). 
The formalism allows to consider a separate time dependence for $\delta\phi(t)$ and $m^2(t)$, so we will keep it unspecified as much as possible.  

\begin{figure}[t]
      \includegraphics[width=0.45\textwidth,angle=0]{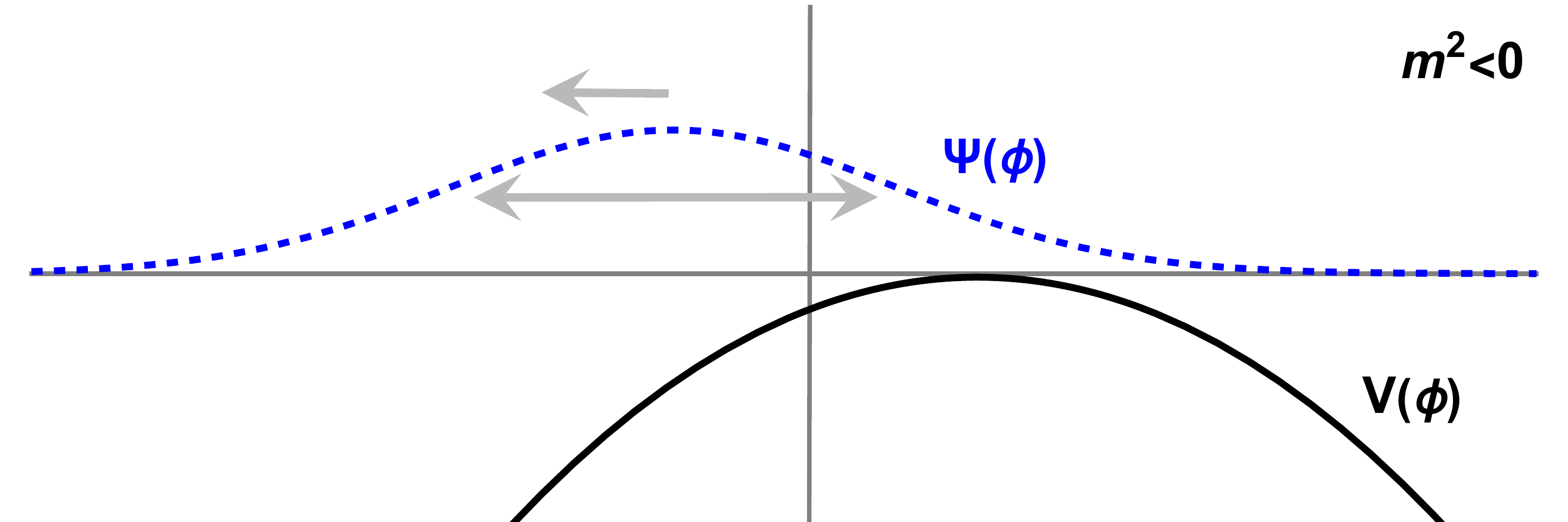}
 \caption{
Sketch of the wavefunctional for the field $\Psi(\phi)$ (dashed blue) and the potential $V(\phi)$ (solid black) as a function of the field $\phi$ after the $\mathbb{Z}_2$ symmetry breaking transition. Generically, after $m^2(t)$ changes sign, $\Psi(\phi)$ is off-center with respect to the maximum of the potential. (This is indeed the case when the bias arises from a constant tilt, see the main text.) The wavefunction then evolves in time by spreading and rolling, as indicated by the arrows. 
The DWs are field configurations that interpolate between the two sides of the maximum.
In practice, we find ourselves in the standard notion of a population bias, with an off-center initial probability distribution.  }
\label{sketch}
\end{figure}

The simplest example of a constant tilt in the potential, $\mu^3 \phi$,  translates to $\delta\phi(t)=-\mu^3/m^2(t)$.  Thus, the instantaneous minimum/maximum of the potential moves to $-\infty$ and reemerges from $+\infty$, for a transient non adiabatic regime. For $m^2(t)$ approaching a negative constant value, the evolution goes back to adiabatic at some point and to a good approximation (if the quench is fast enough) the initial wavefunctional for the field is off-center with respect to the maximum in the potential. In other words, the sudden quench approximation in the presence of a bias reduces to initiating the evolution as illustrated in Fig.~\ref{sketch}.

We postpone a more general study for future work, but here we limit ourselves to this case. As the picture suggests, we are going to capture a bias of the {\em population} type.
Indeed, since in our description the stabilization terms do not appear, the pressure difference $\Delta V$ will  not enter at any stage in the computation. 
In other words, in this paper we will only be interested in the so-called {\it spinodal decomposition} or {\it spinodal instability} phase, where the homogeneous $\phi=0$ vacuum phase splits into different domains with $\phi<0$ and $\phi>0$ separated by domain walls (kinks and antikinks) with arbitrary positions and relative velocities. During this phase, the exact vacuum structure is irrelevant as the field fluctuations aren't yet large enough to `feel' the presence of the stabilizing $\lambda\phi^4$ interaction. This corresponds to the weak-coupling, fast phase transition limit~\cite{Mukhopadhyay:2020gwc}.

The sketch of how the computation proceeds is as follows. The starting point is the quadratic theory \eqref{bias} defined by two functions of time $m^2(t)$, $\delta\phi(t)$. We assume periodic boundary conditions and discretize the spatial coordinate. Effectively, this truncates to a finite number of harmonic oscillators, which allows for a treatment in the Schr\"odinger picture. Since we work at quadratic level, time evolution only amounts to keeping track of how the  Gaussian wavefunctional evolves. 
Following \cite{Mukhopadhyay:2020xmy,Mukhopadhyay:2020gwc}, we identify a kink number density operator, we  extend it to area density for DWs, and simply evaluate its expectation value.

The rest of this article is organized as follows. We review in Sec.~\ref{sec:kinkDiff} a close analogue to DW networks, an ensemble of kinks and antikinks in $1+1$ dimensions, which even in the classical point-particle approximation displays diffusive scaling.
We discuss the quantum version of the problem in  $1+1$ dimensions for kinks  in Sec.~\ref{sec:quantKink}, and  for (precursor) DWs in $d+1$ dimensions in Sec.~\ref{sec:dws}.
We present the effective description of the precursor wall network dynamics in the VOS `one-scale' model language in \ref{sec:VOS}. We briefly discuss our results  in Sec.~\ref{discussion}.

\section{Kink diffusion in $1+1$ dimensions}
\label{sec:kinkDiff}

Kinks in $1+1$ dimensions differ qualitatively from DWs because they aren't extended objects and so they can only behave as massive point particles. As such, a kink/antikink ensemble (or `plasma') must not follow the self-similar scaling common to true DWs. Instead, it falls under a scaling regime of diffusive type. In the following we overview how this happens.  
Let us emphasize that in this Section we consider stabilized kinks (as opposed to kink precursors). Therefore we are having in mind some nonlinear potential for the scalar, the details of which are unimportant beyond providing the kinks as semiclassical solutions that interpolate between the two degenerate minima. For simplicity we stick to a $\mathbb{Z}_2$ invariant model.

Both kinks and antikinks reduce to point particles of the same mass (by $\mathbb{Z}_2$ symmetry). 
At large distances and late times, (anti-)kinks are only characterized by the mass and the point particle approximation must capture well the dynamics. This is the equivalent of the Nambu-Goto description in $1+1$. 
Moreover, they have two more important properties: i) they annihilate when they meet each other (also known as ballistic annihilation), and ii) the $\mathbb{Z}_2$ symmetry breaking transition creates an alternating array of kinks and antikinks. To a very good approximation, the kink/antikink interaction is short range, so they only annihilate when they step on each other. (This holds in the absence of a pressure bias, which would introduce a long range constant force.) 

\begin{figure}[t]
      \includegraphics[width=0.45\textwidth,angle=0]{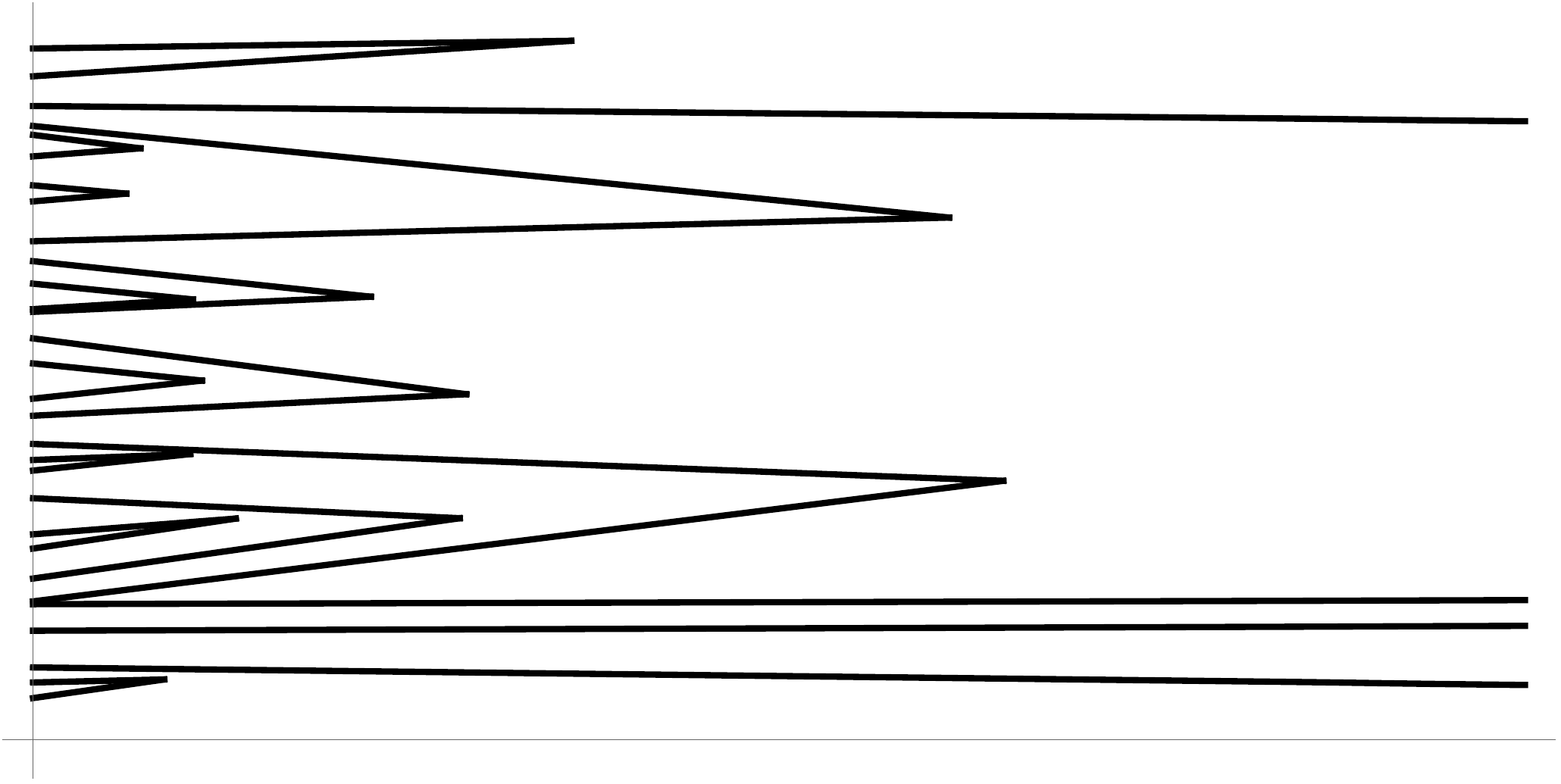}
 \caption{Evolution in time (horizontal axis) of the `kink plasma' -- a distribution of kinks and anti-kinks in 1 space direction (vertical axis), with random initial separations and velocities in a representative realization. Consecutive lines correspond to alternating kinks/anti-kinks. In the point-particle limit, kinks and antikinks annihilate when they meet each other.}
\label{ballistics}
\end{figure}

The system then behaves similarly to an `ionised' plasma, of kinks and antikinks. 
The time evolution is almost trivial since the kink motions are ballistic. One only needs to keep track of when kinks-antikinks collide, and disappear thereafter. See {\em e.g.} Fig.~\ref{ballistics} for an illustrative example of the evolution of the kink-antikink plasma in this approximation.

This kind of problem appeared in the chemistry literature \cite{DecayKinetics,BallisticAnnihilation}, with a focus on whether self-similar evolution and `universal' exponents emerge. The outcome is that the late time behaviour is sensitive to the initial state, the initial distribution of velocities $v$ and inter-particle separations $\Delta x$. This can be somewhat visualized already in Fig.~\ref{ballistics}: pairs that survive for a long time must have small relative velocities. Then, a larger abundance of small velocities in the initial state must translate into a slower dilution.

We can attempt to describe the statistical properties of the kink-antikink plasma with a Boltzmann equation. 
In the simplest description, one keeps track of the kink number density $n_{K}(t)$ and the (root-mean-squared) typical kink velocity $v(t)$.

This is very much parallel to the so-called `velocity dependent one-scale' (VOS) model, developed for cosmic strings \cite{Vilenkin:2000jqa} and later for DW networks \cite{Avelino:2005kn}, to which we return in Sec.\ref{sec:VOS}. 
Then, the annihilation process is expected to be captured by a simple `collision term'
\be\label{boltz}
\dot n_{\rm K} =  -  \sigma \, v  \,  n_{\rm K}^2 
\ee
where $\sigma$ measures the scattering cross section for annihilation (which is dimensionless in $1+1$).  

If the system admits a self-similar solution one expects both $n$ and $v$ to scale as power laws. 
Another way to look at \eqref{boltz} is that the collision rate $\Gamma$ that keeps the plasma `equilibrated' (rather, in a steady-state) is $\Gamma\equiv-\dot n_{\rm K}\,/\,n_{\rm K} =  \sigma \, v \,  n_{\rm K}$.
In a scaling regime, it is reasonable to assume that $\dot v / v$ is also proportional to $\Gamma$. This then provides an evolution equation for $v$,
\be\label{v}
\dot v = -  \kappa \, n_{\rm K} \, v^2 
\ee
with $\kappa$ another dimensionless constant.

By construction, \eqref{boltz} and \eqref{v} must contain scaling solutions. They are parametized by the two constants: the value of $n_{\rm K}\, v\, t$, which must be constant by dimensional analysis	\cite{DecayKinetics}; and the exponent $\mu$, entering as 
$$
n_{\rm K}\sim t^{-\mu} \,, \qquad v\sim t^{\mu-1} ~.
$$
For general $\sigma$, $\kappa$ one indeed finds nontrivial scaling solutions with
$$
\mu = \frac{\sigma}{\kappa+\sigma}~.
$$
The non-universality found in annihilation kinetics \cite{DecayKinetics,BallisticAnnihilation} is the statement that the values of $\sigma$ and $\kappa$ depend on the initial state.  
For $\kappa=\sigma$, Eqns. \eqref{boltz} and \eqref{v} give the diffusive scaling where the separation between pairs, $L\equiv 1/n_{\rm K}$, exhibits Brownian-motion growth
$$
L \sim t^{1/2}~,
$$
typical of diffusion processes. 

Interestingly, this is precisely the behaviour \eqref{nK1+1} found for precursor kinks in $1+1$ using QFT methods \cite{Mukhopadhyay:2020xmy,Mukhopadhyay:2020gwc}. Note that precursor kinks do not necessarily move ballistically. However they also behave like heavy localized objects moving inertially. For this reason, and taking also into account that for precursor kinks the initial condition is fixed by the physical vacuum state of the quantum field \cite{Mukhopadhyay:2020xmy,Mukhopadhyay:2020gwc},  it is not so surprising that the precursor kinks don't differ that much from stabilized kinks and that they actually reduce to a particular choice of $\kappa$, $\sigma$. 
For real DWs (kinks in $d+1$ dimensions with $d>1$) instead we expect that stabilized walls and precursor walls behave differently, see Sec.~\ref{sec:VOS}.

As a final comment, we just remark that at least qualitatively the effect a (pressure) bias is also easy to visualize for the kink-antink ensemble. Indeed, this introduces a constant proper acceleration $a$ that pushes kinks towards antikinks (so that the true vacuum occupies more volume). One expects that this is more efficient than just by diffusion when $\Gamma \lesssim a$.
Clearly, this has catastrophic consequences. At most, after a time of order $1/a$ the velocities become relativistic and the encounter time then is only dictated by the kink-antikink separation. It is clear then that the (exponentially suppressed) abundance of kinks at late times strongly depends on the distribution of separations near the annihilation time $t_{ann}$. We expect that this feature holds for DWs in cosmology too. That is, that the abundance of DWs after annihilation be set by the distribution of closed DWs larger than the Hubble length at $t_{ann}$.

\section{Kink precursors and scaling from free QFT }
\label{sec:quantKink}

We consider a scalar field $\phi(t,x)$ in $1+1$ dimensions undergoing a quantum phase transition. 
Assuming a potential of the form \eqref{bias}, the model is parameterized by 2 functions: $m^2(t)$ and $\delta\phi(t)$.
To set the stage, we assume that initially the scalar potential is $\mathbb{Z}_2$-symmetric and has a unique vacuum at $\phi=0$ (where the mass of $\phi$ is $m_0$). As time goes  on, the potential flips (under the effect of some external time-varying parameter for instance) and it acquires two local minima. To keep it quite general, we allow the local maximum of this new potential (where the tachyonic mass of $\phi$ is assumed to be of magnitude $m_0$) to be shifted by some amount $\delta\phi_0$. Without loss of generality we also assume that the potential vanishes at this point. The $\mathbb{Z}_2$-symmetry is thus both explicitly and spontaneously broken during the phase transition.

This stage of the phase transition can thus described by the quadratic Lagrangian
\be
\mathcal{L} = \int dx \left[\half \dot{\phi}^2-\half \phi^\prime{}^2 - \half m^2(t) (\phi-\delta\phi (t))^2\right] \,,
\label{model}
\ee
where $m^2(t)$ and $\delta\phi(t)$ are functions of time only that verify, $m^2=m_0^2$ and $\delta\phi=0$ at $t\to-\infty$, while $m^2=-m_0^2$ and $\delta\phi =\delta\phi_0$ at $t\to+\infty$. This can be achieved for instance by taking $m^2(t) = - m_0^2 \tanh \left ( t/\tau_m \right )$ and $\delta\phi(t)=\delta\phi_0\left[1+\tanh\left((t-t_s)/\tau_s\right)\right]$ where the $t_s$, $\tau_m$ and $\tau_s$ are time scales that parametrize the details of the quench 
(we shall focus on the  instantaneous quench limit $\tau_{m,\,s}\to0$ with $t_s=0$ later on).
Dotted and primed quantities respectively denote time and space derivatives.

In this context, the natural definition of kinks  are as the zeros of $\phi-\delta\phi$. (We work at quadratic level, so keep in mind that these are kink precursors really.) 
Our task then is to estimate the evolution of the average kink number density over the course of the phase transition. Since the model simply corresponds to a free quantum scalar field evolving in a time dependent homogeneous background, we will proceed by fully solving for its quantum dynamics in the Schr\"odinger picture. We will closely follow the methods of~\cite{Mukhopadhyay:2020xmy,Mukhopadhyay:2020gwc}, by solving for the time-dependent wavefunctional describing the quantum state of the field $\phi$ in the Schr\"odinger picture and using this to semi-analytically compute the quantum average of a properly defined kink number density operator.

\subsection{Dynamics in the Schr\"odinger picture}

We start by compactifying the model {\it i.e.} imposing periodic boundary conditions at $x=0$ and $x=L$. After an integration by parts, \eqref{model} can thus be recast as
\be
\mathcal{L}=\int dx\left[\half \dot{\phi}^2+\half \phi\,\phi^{\prime\prime} - \half m^2(t) (\phi-\delta\phi (t))^2\right]\,.
\label{modelibp}
\ee
We then discretize it on a finite lattice made up of $N$ evenly spaced points labelled by an index $\mathscr{i}$ running from $1$ to $N$. The lattice spacing $L/N$ is denoted by $a$. If we define the discretized field values $\phi_\mathscr{i} \equiv \phi(t,\mathscr{i} a)$ and replace the second derivative by its lowest order central finite difference approximation,
\be
\phi_\mathscr{i}^{\prime\prime} \to \frac{\phi_{\mathscr{i}-1}-2\phi_\mathscr{i}+\phi_{\mathscr{i}+1}}{a^2}\,,
\ee
we can rewrite~\eqref{modelibp} as
\ba
\mathcal{L}&=&\sum_{\mathscr{i}=1}^N a\left[\half\dot{\phi}_\mathscr{i}^2+\frac{1}{2a^2}\phi_\mathscr{i}(\phi_{\mathscr{i}-1}-2\phi_\mathscr{i}+\phi_{\mathscr{i}+1})\right.\nn\\
&&\hspace{3cm}\left.-\half m^2(t) (\phi_\mathscr{i}-\delta\phi (t))^2\right]\,.
\label{modeldisc}
\ea
Notice that $\phi_{N+1}=\phi_1$ and $\phi_0=\phi_N$ by virtue of the periodicity of the lattice. 

We can give a more compact expression of this Lagrangian by assembling the discretized field values in a column vector $\bm{\phi}\equiv (\phi_1, \phi_2, \dots , \phi_N)^T$ and defining the matrix $\bm{\Omega}^2(t)$ by
\be
[\bm{\Omega}^2(t)]_{\mathscr{i}\mathscr{j}} = 
\begin{cases}
+{2}/{a^2}+m^2(t)\,,& \mathscr{i}=\mathscr{j}\\
-{1}/{a^2}\,,& \mathscr{i}=\mathscr{j}\pm1\ (\text{mod}\ N)\\
0\,,&\text{otherwise}\,.
\end{cases}
\ee
(Here and henceforth, bold quantities denote vectors and matrices while ${}^T$ denotes matrix transposition.) With these definitions and conventions Eq.~\eqref{modeldisc} reads
\be
\mathcal{L}= \frac{a}{2}\dot{\bm{\phi}}^T.\dot{\bm{\phi}}-\frac{a}{2}(\bm{\phi}-\delta\phi(t)\bm{1})^T.\bm{\Omega_2}(t).(\bm{\phi}-\delta\phi(t)\bm{1})\,,
\label{disc_lag}
\ee
where we have introduced the `vector of ones' $\bm{1}\equiv(1, 1, \dots ,1)^T$ and used the property $\bm{\Omega}^2(t).\bm{1}=m^2(t)\bm{1}$ ({\it i.e.}, $\bm{1}$ is an eigenvector associated to the eigenvalue $m^2(t)$).

Since the Lagrangian is quadratic, and assuming that at $t=t_0\ll -m_0^{-1}$ the field starts in its (gaussian) quantum vacuum, we know that the state will remain gaussian during time evolution. A good ansatz for the Schr\"odinger wavefunctional describing the state of the system at time $t$ is therefore
\be
\Psi(\bm{\phi},t)=\mathcal{N}(t)\exp\left[ia\bm{D}(t)^T.\bm{\phi}+\frac{ia}{2}\bm{\phi}^T.\bm{M}(t).\bm{\phi}\right]\,,
\label{ansatz}
\ee
where $\bm{M}(t)$ is an $N\times N$ complex symmetric matrix, $\bm{D}(t)$ a complex $N$-component column vector and $\mathcal{N}(t)$ a normalization factor. The requirement that the field $\phi$ be in its vacuum long before the phase transition imposes $\bm{M}(t_0)=i\bm{\Omega}^2(t_0)^{1/2}$, $\bm{D}(t_0)=0$, and $\mathcal{N}(t_0)=(a/\pi)^{N/4}\det(\bm{\Omega}^2(t_0))^{1/8}$. Here the matrix powers are unambiguously defined since the matrix $\bm{\Omega}^2(t_0)$ is symmetric positive definite.

Now, $\Psi$ verifies the Schr\"odinger equation
\ba
i\frac{\partial \Psi}{\partial t}&=&-\frac{1}{2a}\sum_{\mathscr{i}=1}^N\frac{\partial^2\Psi}{\partial\phi_\mathscr{i}^2}\nn\\
&&+\frac{a}{2}(\bm{\phi}-\delta\phi(t)\bm{1})^T.\bm{\Omega}^2(t).(\bm{\phi}-\delta\phi(t)\bm{1})\,\Psi\,,
\label{schrodinger}
\ea
where we have set $\hbar=1$. Plugging~\eqref{ansatz} into~\eqref{schrodinger} and identifying the quadratic terms in $\phi_\mathscr{i}\phi_\mathscr{j}$ yields the matrix dfiferential equation
\be
\dot{\bm{M}}+\bm{M}^2+\bm{\Omega}^2(t)=0\,.
\ee
It is easy to convince oneself that the unique solution to this equation subject to the specified initial conditions is symmetric for all times. Analogously, identifying the linear terms in $\phi_\mathscr{i}$ (and using the fact that $\bm{1}$ is an eigenvector of $\bm{\Omega}^2$) yields
\be
\dot{\bm{D}}+\bm{M}.\bm{D}-m^2(t)\delta\phi(t)\bm{1}=0\,.
\ee
The normalization factor is also easy to compute as a function of $\bm{M}$ but we will not need it explicitly. It turns out that one can solve for $\bm{M}$ and $\bm{D}$ in terms of $N$ complex functions of time.

We now introduce the complex mode functions $c_n(t)$ (with $n$ running from $0$ to $N-1$) verifying
\be
{\ddot c}_n + \left [ \frac{4}{a^2}\sin^2 \left (\frac{\pi n}{N} \right ) + m^2(t) \right ] c_n = 0\,,
\label{ckeq}
\ee
and with initial conditions 
\ba
c_n (t_0) &=& 
\frac{1}{\sqrt{2a}} 
\left [ \frac{4}{a^2}\sin^2 \left ( \frac{\pi n}{N}\right ) + m^2(t_0) \right ]^{-1/4}\,,
\label{ckt0}\\
{\dot c}_n (t_0) &=& \frac{i}{\sqrt{2a}} 
\left [ \frac{4}{a^2}\sin^2 \left (\frac{\pi n}{N} \right ) + m^2(t_0) \right ]^{1/4}\,.
\label{dotckt0}
\ea
Then it is straightforward to verify that
\ba
[\bm{M}(t)]_{\mathscr{i}\mathscr{j}}&=&\frac{1}{N}\sum_{n=0}^{N-1} c_n(t)^{-1}\dot{c}_n(t)e^{2i\pi n(\mathscr{i}-\mathscr{j})/N}\label{Mexp}\\
&=&\frac{1}{N}\sum_{n=0}^{N-1} c_n(t)^{-1}\dot{c}_n(t)\cos(2\pi n(\mathscr{i}-\mathscr{j})/N)\,,\nn
\ea
where we have used the fact that $c_{n}(t)=c_{N-n}(t)$ for $1\leq n\leq N-1$ to write $\bm{M}$ in manifestly symmetric form. Similarly we can see that
\be
\bm{D}(t)=c_0(t)^{-1}\int_{t_0}^t ds\, m^2(s)\,\delta\phi(s) c_0(s) \bm{1}\,.
\label{Dexp}
\ee

We are finally in a position to write the probability density functional of the field configuration $\bm{\phi}$,
$\mathcal{P}(\bm{\phi},t)\equiv |\Psi(\bm{\phi},t)|^2$,
in terms of the $c_n(t)$. We have
\be
\mathcal{P}(\bm{\phi},t)=\frac{1}{\sqrt{\det(2\pi \bm{K})}}e^{-(\bm{\phi}-\Delta(t)\bm{1})^T.\bm{K}(t)^{-1}.(\bm{\phi}-\Delta(t)\bm{1})/2}\,,
\label{probdensity1d}
\ee
where
\be
[\bm{K}(t)]_{\mathscr{i}\mathscr{j}}=\frac{1}{N}\sum_{n=0}^{N-1}|c_n(t)|^{2}\cos(2\pi n(\mathscr{i}-\mathscr{j})/N)
\ee
is the spatial correlation function between two lattice points,
and
\ba
\Delta(t) &=& ia\int_{t_0}^t ds\,m^2(s)\,\delta\phi(s)\nn\\
&&\hspace{1.2cm}\times\left(c_0(t)^*c_0(s)-c_0(t)c_0(s)^*\right)
\ea
represents the (homogeneous) shift of the vacuum expectation value of $\phi$ triggered by the potential bias $\delta\phi$. To derive these expressions we have used~\eqref{Mexp} and~\eqref{Dexp} as well as the fact that
$c_n(t)^*\dot{c}_n(t)-c_n(t)\dot{c}_n(t)^*$ is a conserved quantity equal to $i/a$ for all $n$. As a sanity check, it is easy to verify that for constant positive $m^2(t)=m_0^2$ and slowly varying $\delta\phi(t)$, $\Delta(t)\approx\delta\phi(t)$.

We now have all the necessary tools at our disposal to evaluate the kink number density.

\subsection{Kink number density}
\label{numberdensity}

We will follow the method in Refs.~\cite{Mukhopadhyay:2020xmy,Mukhopadhyay:2020gwc} and look for kinks and antikinks among zeros of $\phi-\delta\phi(t)$. Of course some of those will be quantum fluctuations and there will be a vast overcounting but we will turn to that problem later on. For the time being we define the quantum operator
\ba
\hat{n}_{\rm Z}&\equiv&\frac{1}{L}\sum_{\mathscr{i}=1}^N\frac{1}{4}\left[\rm{sgn}(\hat{\phi}_\mathscr{i}-\delta\phi(t))-\rm{sgn}(\hat{\phi}_{\mathscr{i}+1}-\delta\phi(t))\right]^2\nn\\
&=&\frac{N}{2L}-\frac{1}{2L}\sum_{\mathscr{i}=1}^N \rm{sgn}\left((\hat{\phi}_\mathscr{i}-\delta\phi)(\hat{\phi}_{\mathscr{i}+1}-\delta\phi)\right)\,,
\ea
where $\rm{sgn}$ denotes the signum function. This operator counts the number density of $\phi-\delta\phi(t)$ sign changes between two consecutive lattice points thus providing a lower estimate for the number density of zeros of this quantity. Of course this estimate will change as the lattice gets finer (or in other words as $a$ becomes smaller) and, given the fundamentally quantum nature of the problem, may potentially diverge in the continuum limit. We will come back to this problem and for the time being will simply disregard any subtleties related to the coarseness of the lattice. In fact we want to calculate the quantum average of this operator in the state whose wavefunctional we computed previously. This reads
\be
\langle\hat{n}_{\rm Z}\rangle=\frac{N}{2L}-\frac{1}{2L}\sum_{\mathscr{i}=1}^N\left\langle \rm{sgn}\left((\hat{\phi}_\mathscr{i}-\delta\phi)(\hat{\phi}_{\mathscr{i}+1}-\delta\phi)\right)\right\rangle\,,
\label{nz1}
\ee
where, after a shift in the integration variables,
\ba
&&\left\langle \rm{sgn}\left((\hat{\phi}_\mathscr{i}-\delta\phi)(\hat{\phi}_{\mathscr{i}+1}-\delta\phi)\right)\right\rangle=\nn\\
&&\hspace{0.5cm}\int d\phi_1\dots d\phi_N \rm{sgn}\left(\phi_\mathscr{i}\phi_{\mathscr{i}+1}\right)\mathcal{P}(\bm{\phi}+\delta\phi(t)\mathbf{1},t)\,.
\ea
Now, since the matrix $\bm{K}^{-1}$ is circulant {\it i.e.} its coefficients $[\bm{K}^{-1}]_{\mathscr{i}\mathscr{j}}$ only depend on $\mathscr{i}-\mathscr{j}$, it is easy to see that $\left\langle \rm{sgn}\left((\hat{\phi}_\mathscr{i}-\delta\phi)(\hat{\phi}_{\mathscr{i}+1}-\delta\phi)\right)\right\rangle=\left\langle \rm{sgn}\left((\hat{\phi}_1-\delta\phi)(\hat{\phi}_{2}-\delta\phi)\right)\right\rangle$. (This is the algebraic signature of the fact that the background is translationally invariant.) If we define the four quadrants of the $(\phi_1, \phi_2)$ plane in the conventional way, Eq.~\ref{nz1} reduces to
\be
\langle\hat{n}_{\rm Z}\rangle=\frac{N}{2L}\left[1+\sum_{Q=1}^4(-1)^{Q}I_Q\right]\,,
\label{nz2}
\ee
where
\be
I_Q\equiv\iint_{Q\ \rm{quadrant}}\hspace{-0.5cm}d\phi_1d\phi_2\tilde{\mathcal{P}}(\phi_1+\delta\phi(t),\phi_2+\delta\phi(t))
\ee
and
\be
\tilde{\mathcal{P}}(\phi_1,\phi_2)\equiv \int d\phi_3\dots d\phi_N\mathcal{P}(\bm{\phi},t)
\ee
is the marginal probability density of $\phi_1$ and $\phi_2$. We can easily compute this quantity by using a well-known property of multi-variate normal distributions such as $\mathcal{P}$ which states that in order to obtain the marginal distribution over a subset of variables, one simply needs to drop the variables that are intregrated out from the covariance matrix and the mean vector (see for instance \cite{bishop2013pattern}). 
In our case the mean vector is zero and the marginal covariance matrix reduces to the $2\times 2$ upper left block of $\bm{K}$ or $\bm{K}_{2\times 2}\equiv\alpha \bm{I}_{2} +\beta\bm{\sigma}_1$ where
\ba
\alpha(t)&\equiv& [\bm{K}(t)]_{11}=\frac{1}{N}\sum_{n=0}^{N-1}|c_n(t)|^{2}\,,\\
\beta(t)&\equiv&[\bm{K}(t)]_{12}=\frac{1}{N}\sum_{n=0}^{N-1}|c_n(t)|^{2}\cos(2\pi n/N)\,.
\ea
With these notations we obtain
\ba
\tilde{\mathcal{P}}(\phi_1,\phi_2)&=&\frac{1}{2\pi\sqrt{\alpha^2-\beta^2}}\exp\biggl[-\frac{1}{2(\alpha^2-\beta^2)}\Big(\alpha(\phi_1-\Delta)^2\nn\\
&&\hspace{0.8cm}-2\beta(\phi_1-\Delta)(\phi_2-\Delta)+\alpha(\phi_2-\Delta)^2\Big)\biggr]\,.\nn
\ea

Noticing that $\sum_{Q=1}^4I_Q=1$ and that $I_2=I_4$ we can now rewrite~\eqref{nz2} in the more explicit manner,
\begin{widetext}
\be
\langle\hat{n}_{\rm Z}\rangle = \frac{2N}{L}I_2=\frac{N}{L\sqrt{\pi}}\int_0^{\infty}dx\,e^{-x^2}\left[{\rm erf}\left(x\sqrt{\frac{\alpha-\beta}{\alpha+\beta}}-\frac{\delta\phi-\Delta}{\sqrt{\alpha+\beta}}\right)+{\rm erf}\left(x\sqrt{\frac{\alpha-\beta}{\alpha+\beta}}+\frac{\delta\phi-\Delta}{\sqrt{\alpha+\beta}}\right)\right]\,,
\label{nz3}
\ee
\end{widetext}
where ${\rm erf}(z)=2/\sqrt{\pi}\int_0^zdt\,e^{-t^2}$ is the standard error function.
As stated at the beginning of this section, if we simply use this formula as is, we will vastly overcount the number density of kinks on the lattice. This is due to small fluctuations of the quantum field that are always present (whether there is a phase transition or not) and that should be disregarded. Following Refs.~\cite{Mukhopadhyay:2020xmy,Mukhopadhyay:2020gwc} we notice that such spurious zeros of the field that do not correspond to a kink or antikink are due to the presence of oscillatory modes in the expressions for $\alpha$ and $\beta$. We are thus led to define $\bar{\alpha}$ and $\bar{\beta}$ by restricting the sums to those modes $c_n$ that are unstable {\it i.e.} with $n$ such that $4\sin^2(\pi n/N)/a^2+m^2(t)\leq 0$,

\ba
\bar{\alpha}(t)&\equiv& \frac{1}{N}\left[|c_0(t)|^2+2\sum_{n=1}^{n_c(t)}|c_n(t)|^{2}\right]\!\!,\\
\bar{\beta}(t)&\equiv&\frac{1}{N}\left[|c_0(t)|^2+2\sum_{n=1}^{n_c(t)}|c_n(t)|^{2}\cos(2\pi n/N)\right]\!\!,
\ea
where $n_c(t)\equiv\lfloor N\sin^{-1}(a \sqrt{-m^2(t)}/2)/\pi\rfloor$ and $\lfloor\rfloor$ denotes the integer part function. (When there are no unstable modes, $\bar{\alpha}$ and $\bar{\beta}$ are understood to vanish by convention.) Then the average number density of kinks $n_{\rm K}$ can be computed from $\langle\hat{n}_{\rm Z}\rangle$ by replacing $\alpha$ and $\beta$ by $\bar{\alpha}$ and $\bar{\beta}$ in~\eqref{nz3},
\ba
n_{\rm K} &=& \frac{N}{L\sqrt{\pi}}\int_0^{\infty}dx\,e^{-x^2}\left[{\rm erf}\left(x\sqrt{\frac{\bar{\alpha}-\bar{\beta}}{\bar{\alpha}+\bar{\beta}}}-\frac{\delta\phi-\Delta}{\sqrt{\bar{\alpha}+\bar{\beta}}}\right)\right.\nn\\
&&\hspace{1.5cm}\left.+{\rm erf}\left(x\sqrt{\frac{\bar{\alpha}-\bar{\beta}}{\bar{\alpha}+\bar{\beta}}}+\frac{\delta\phi-\Delta}{\sqrt{\bar{\alpha}+\bar{\beta}}}\right)\right]\,.
\label{nk1}
\ea

We can already notice that for zero bias, $\delta\phi(t)=\Delta(t)=0$ and we recover the result of Refs.~\cite{Mukhopadhyay:2020xmy,Mukhopadhyay:2020gwc},
\be
\left.n_{\rm K}\right|_{\rm no\, bias}=\frac{2N}{\pi L}\tan^{-1}\left(\sqrt{\frac{\bar{\alpha}-\bar{\beta}}{\bar{\alpha}+\bar{\beta}}}\right)=\frac{N}{\pi L}\cos^{-1}\left(\frac{\bar{\beta}}{\bar{\alpha}}\right)\,.
\label{nknobias}
\ee

Before going any further it will be interesting to see how~\eqref{nk1} and~\eqref{nknobias} simplify in the continuum limit {\it i.e.} when $N\to\infty$ at fixed $L$. For this we notice that the quantity
\be
\sqrt{\frac{\bar{\alpha}-\bar{\beta}}{\bar{\alpha}+\bar{\beta}}}=\tan\left(\frac{\pi L }{2N}\left.n_{\rm K}\right|_{\rm no\, bias}\right)\approx\frac{\pi L }{2N}\left.n_{\rm K}\right|_{\rm no\, bias}
\ee
vanishes in this limit since $\left.n_{\rm K}\right|_{\rm no\, bias}$ is a physical quantity that has a well-defined, $N$ independent, finite limit for all times~\cite{Mukhopadhyay:2020xmy,Mukhopadhyay:2020gwc}. 
Moreover, the integral in~\eqref{nk1} is dominated by values of the integrand close to $x=0$ (because of the $e^{-x^2}$ factor). We can therefore Taylor expand the error functions in powers of the vanishingly small quantity $x\sqrt{(\bar{\alpha}-\bar{\beta})/(\bar{\alpha}+\bar{\beta})}$ to obtain
\be
n_{\rm K}= \left.n_{\rm K}\right|_{\rm no\, bias} \exp\left[-\frac{(\delta\phi-\Delta)^2}{\bar{\alpha}+\bar{\beta}}\right]\,.
\label{nk2}
\ee
It is worth noticing that this expression only depends on $L$ through $\bar{\alpha}$ and $\bar{\beta}$ which are sums of a finite number ($2n_c(t)+1\leq L\sqrt{-m^2(t)}/4$) of terms. As $L\to\infty$ these sums become integrals.

\begin{figure}[t]
      \includegraphics[width=0.45\textwidth,angle=0]{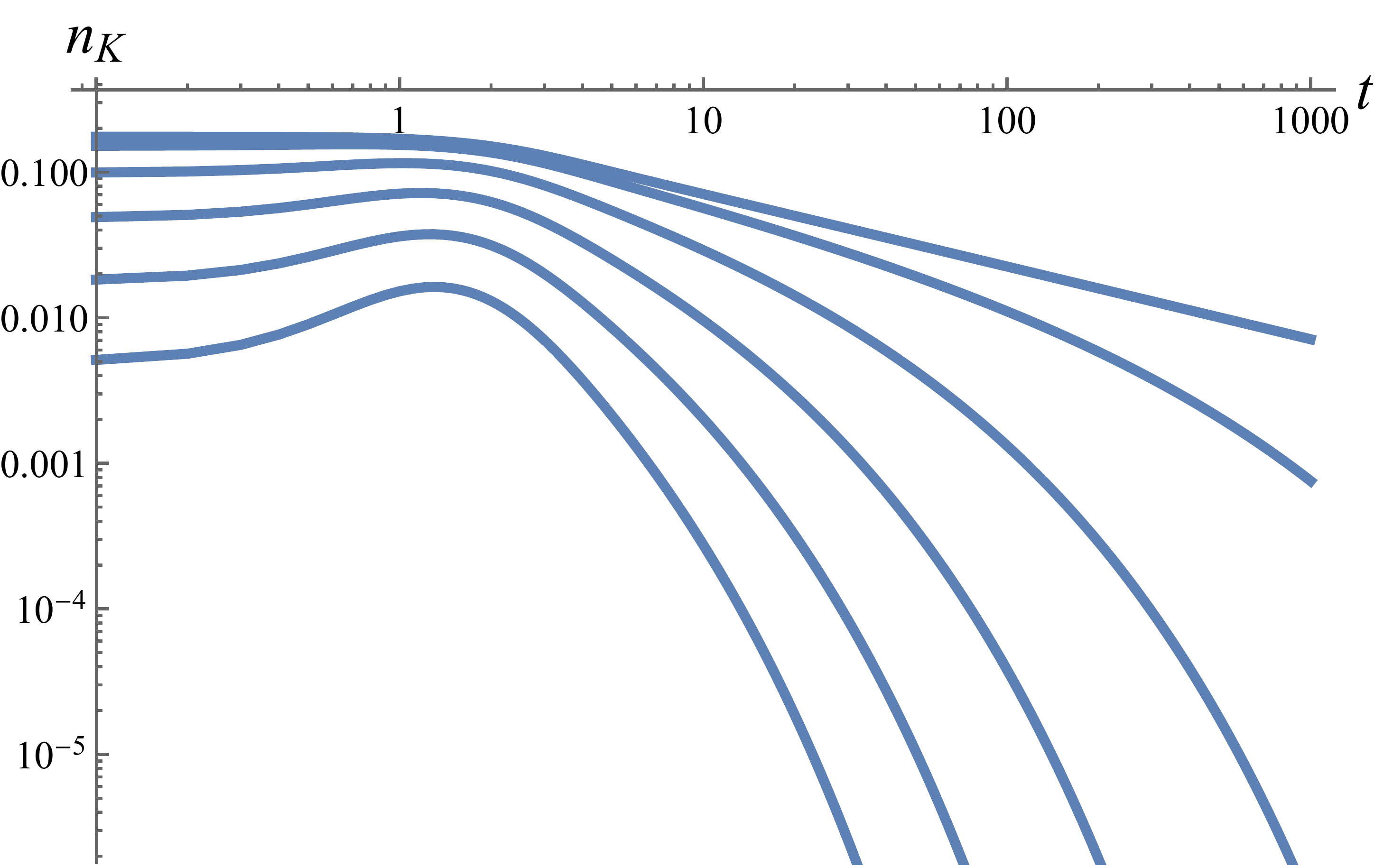}
 \caption{
Kink number density as a function of time in $m_0=1$ units  for different values of the bias $\delta\phi_0$, in $1+1$ dimensions. The topmost curve corresponds to the zero bias case, and $\delta\phi_0$ increases in increments of $0.2$ as we move down to the lower curves. 
  }
\label{resplot0}
\end{figure}

\begin{figure}[t]
      \includegraphics[width=0.45\textwidth,angle=0]{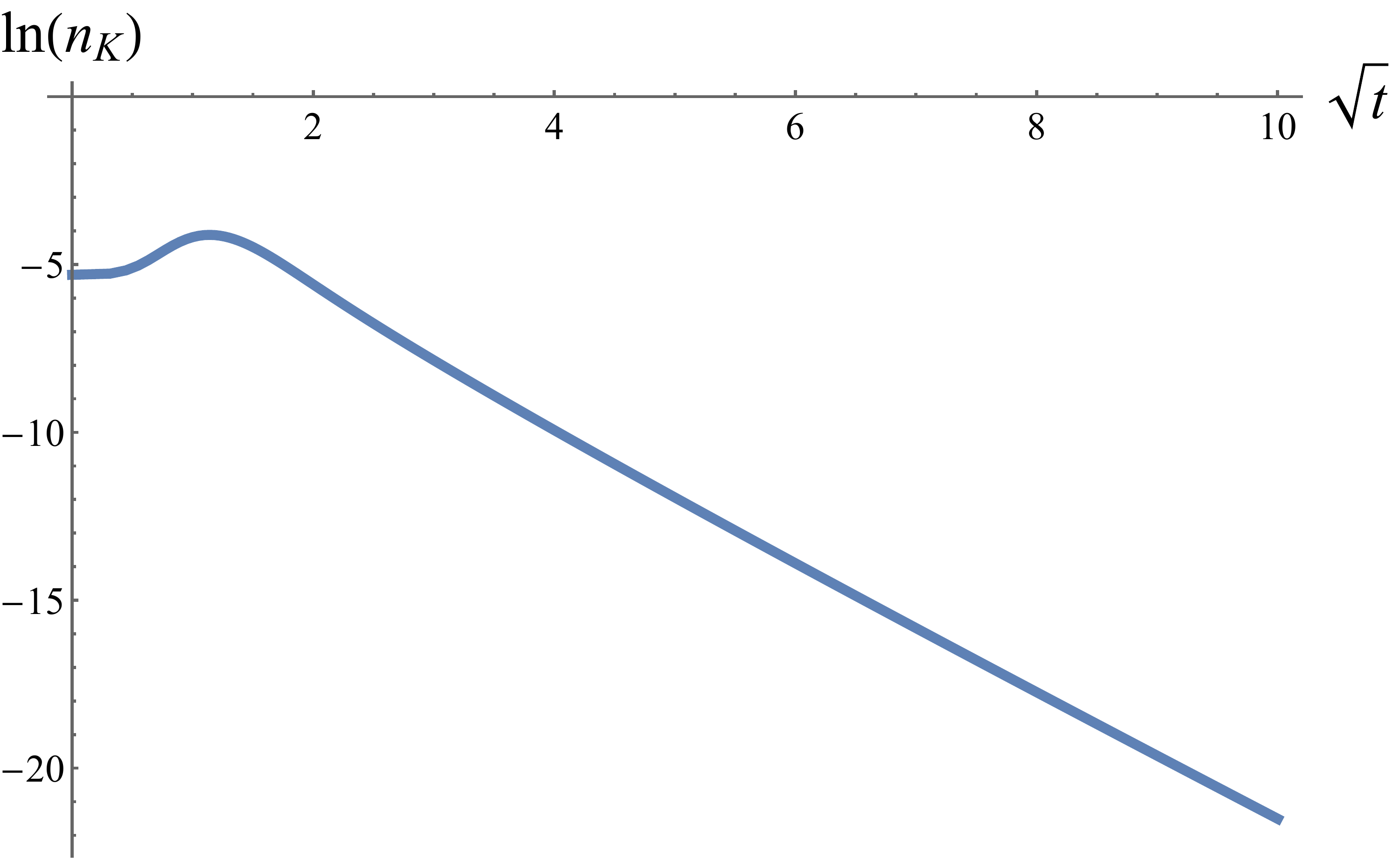}
 \caption{
Plot of the natural logarithm of the kink number density as a function of the square root of time in  $m_0=1$ units and for $\delta\phi_0=1$. 
  }
\label{resplot1}
\end{figure}

We now turn to the evaluation of $n_{\rm K}$ which can be done either semi-analytically in the particular case of a sudden phase transition,
where $m^2(t)=-m_0^2 (2\Theta(t)-1)$ and $\delta\phi(t)=\delta\phi_0\Theta(t)$ ($\Theta$ being the standard Heaviside step function). In this case the mode functions $c_n(t)$ are exactly calculable and, taking first the continuum limit $N\to\infty$ and then the infinite volume limit $L\to\infty$,we obtain
\ba
&&\hspace{-0.4cm}\delta\phi(t)-\Delta(t) = \delta\phi_0\cosh\left[m_0(t-t_s)\right]\Theta(t-t_s)\,,\\ 
&&\hspace{-0.4cm}\bar{\alpha}+\bar{\beta} = \frac{1}{\pi}\bigintsss_0^{m_0} \hspace{-0.4cm}dk\,\left[\frac{m_0^2\cosh\left[2t\sqrt{m_0^2-k^2}\right]-k^2}{(m_0^2-k^2)\sqrt{m_0^2+k^2}}\right]
\label{kint1}
\ea
and $\left.n_{\rm K}\right|_{\rm no\, bias}$ is known from Eq.~(93) of Ref.~\cite{Mukhopadhyay:2020gwc} to be

\ba
&& \frac{1}{\pi}\left\{\bigintsss_{0}^{m_0} \hspace{-0.4cm} dk\, k^2
\left[\frac{m_0^2\cosh\left(2t\sqrt{m_0^2-k^2}\right)-k^2}{(m_0^2-k^2)\sqrt{k^2+m_0^2}}\right]\right\}^{1/2}\nn\\
&&\hspace{0.5cm}\times
\left\{\bigintsss_{0}^{m_0} \hspace{-0.4cm} dk\,
\left[\frac{m_0^2\cosh\left(2t\sqrt{m_0^2-k^2}\right)-k^2}{(m_0^2-k^2)\sqrt{k^2+m_0^2}}\right]\right\}^{-1/2}\,.
\label{kint2}
\ea
Now, plugging these expression into~\eqref{nk2} yields an explicit analytic expression that we plot in Fig.~\ref{resplot0}. As expected we notice that the initial kink number density is orders of magnitude lower than in the zero bias case. This is due to the presence of a potential barrier that suppresses kink formation. A surprising feature of this plot is the small bump appearing around $t\sim m_0^{-1}$ and that decays rapidly afterwards. Technically, this is due to the vacuum expectation value of the field $\Delta(t)$ growing more slowly than the ``width'' $\sim \bar{\alpha}+\bar{\beta}$ of the probability density functional. Physically, it seems that we can interpret this are bubble nucleation in the first stages of evolution, when the potential barrier is still not too high and tunnelling shouldbe not suppressed. The late time behavior is indicative of a faster than power law suppression. In fact, in this limit, $n_{\rm K}/\left.n_{\rm K}\right|_{\rm no\, bias}$  is well fit by $\exp[-1.72\, t^{0.51}]$ in units where $m_0=1$, as can be seen in Fig.~\ref{resplot1}.

We can even recover explicitly the late time behavior of $n_{\rm K}(t)$. Realizing that in the limit $t\gg m_0$ all the $k$ integrals in~\eqref{kint1} and~\eqref{kint2} are dominated by values of $k\ll m_0$ we can replace the integrands with their lowest order $k/m_0$ expansions to obtain
\be
n_{\rm K}(t)\sim
\sqrt{\frac{m_0}{t}}e^{-\delta\phi_0^2\sqrt{\pi m_0 t}}\,.
\ee
This is in good agreement with the above numerically determined coefficients.

\section{Domain wall networks}
\label{sec:dws}

The same method outlined above for kinks in $1+1$ dimensions can be extended rather directly to $d+1$ dimensions. For concreteness we will show explicitly how the extension works in $2+1$ dimensions and then give the general result.

We start with the Lagrangian for the $2+1$ dimensional real scalar field $\phi(t,x,y)$
\ba
\label{highd_lagrangian}
\mathcal{L}&=&\int dx dy\left[ \half (\partial_t\phi)^2-\half (\partial_x\phi)^2-\half (\partial_y\phi)^2 \right.\nn\\
&&\hspace{2cm}\left.- \half m^2(t) (\phi-\delta\phi(t))^2\right]\,,
\ea
where the functions of time $m^2(t)$ and $\delta\phi(t)$ obey the same properties as in the $1+1$ dimensional case. (The field $\phi$ is now dimnensionful however.) This is seen to be a model with a broken $\mathbb{Z}_2$ symmetry that typically would feature domain walls. Next, we compactify space on a 2-torus of area $L^2$ by assuming periodic boundary conditions ($\phi(x+L,y)=\phi(x,y+L)=\phi(x,y)$) and discretize it on a regular square lattice made up of $N^2$ (with lattice spacing $a=L/N$). At each lattice point $(x_{\mathscr{i}},y_{\mathscr{j}}) \equiv (\mathscr{i}a,\mathscr{j}a)$ we define the discretized 
field values $\phi_{\mathscr{i}\mathscr{j}} \equiv \phi (x_{\mathscr{i}},y_{\mathscr{j}})$.

The discretized version of~\eqref{highd_lagrangian} has a form analogous to~\eqref{disc_lag},
\be
\mathcal{L}= \frac{a^2}{2}\dot{\bm{\phi}}^T.\dot{\bm{\phi}}-\frac{a^2}{2}(\bm{\phi}-\delta\phi(t)\bm{1})^T.\bm{\Omega_2}(t).(\bm{\phi}-\delta\phi(t)\bm{1})\,,
\ee
as long as it is understood that any vectors and matrices are now $N^2$ and $N^2\times N^2$ dimensional respectively. For instance, 
\be
{\bm \phi} \equiv (\phi_{11},\phi_{12},...,\phi_{1N},
\phi_{21},...,\phi_{2N},...,\phi_{NN-1},\phi_{NN})^T.\nn
\ee 
and $\bm{\Omega}^2(t)$ is given by
\be
[\bm{\Omega}^2]_{\mathscr{i}\mathscr{j},\mathscr{k}\mathscr{l}} = 
\begin{cases}
+{2}/{a^2}+m^2(t)\,,& \mathscr{i}=\mathscr{k},\mathscr{j}=\mathscr{l}\\
-{1}/{a^2}\,,& \mathscr{i}=\mathscr{k}\pm1, \mathscr{j}=\mathscr{l}\pm1 \\
0\,,&\text{otherwise}\,,\nn
\end{cases}
\ee
where equality relations are understood to modulo $N$. More generally, any $N^2\times N^2$ matrix $\bm{A}$ would be represented by a two-dimensional array of matrix elements $A_{\mathscr{i}\mathscr{j},\mathscr{k}\mathscr{l}}$ arranged in the following way:
\be
\label{fourindexmatrix}
A =
\begin{pmatrix}
A_{11,11}          & A_{11,12} & \cdots & A_{11,1N} & A_{11,21} & A_{11,22} & \cdots            \\
A_{12,11}          & A_{12,12} & \cdots & A_{12,1N} & A_{12,21} & A_{12,22} & \cdots             \\
\vdots  & \vdots  &    & \vdots  & \vdots   & \vdots    \\
A_{1N,11}          & A_{1N,12} & \cdots & A_{1N,1N} & A_{1N,21} & A_{1N,22} & \cdots             \\
A_{21,11}            &   &    &                           \\
A_{22,11}          &   &    &                               \\
  \vdots          &   &    &                                               
\end{pmatrix}
\nn
\ee
The generalization of the results of the previous sections is now straightforward. The functional Schr\"odinger equation again has a Gaussian solution $\Psi(t,\bm{\phi})$ which can be expressed in terms of the two-deimansional mode functions $c_{n,m}(t)$ (with both $n$ and $m$ running from $0$ to $N-1$). These verify
\be
{\ddot c}_{n,m} + \left [ k_{n,m}^2 + m^2(t) \right ] c_{n,m} = 0\,,
\label{ck2deq}
\ee
and with initial conditions 
\ba
c_{\mathscr{n},m} (t_0) &=& 
\frac{1}{a\sqrt{2}} 
\left [ k_{n,m}^2 + m^2(t_0) \right ]^{-1/4}\,,\\
\label{ck2dt0}
{\dot c}_{n,m} (t_0) &=& \frac{i}{a\sqrt{2}} 
\left [k_{n,m}^2 + m^2(t_0) \right ]^{1/4}\,.
\label{dotck2dt0}
\ea
Here we have introduced the discretized momentum 
\be
k_{n,m}=\frac{2}{a}\left\{\sin^2 \left (\frac{\pi n}{N} \right ) + \sin^2 \left (\frac{\pi m}{N} \right )\right\}^{1/2}\,,
\ee
for notational simplicity. (Notice also that the  normalization of the mode functions has been modified with respect to~\eqref{ckt0} and~\eqref{dotckt0} in such a way that they have the same dimension as the field $\phi$ {\it i.e.} 1/2.) The probability density functional $\mathcal{P}(t,\bm{\phi})=|\Psi(t,\bm{\phi})|^2$ is of the same form as~\eqref{probdensity1d} but with
\ba
[\bm{K}(t)]_{\mathscr{i}\mathscr{j},\mathscr{k}\mathscr{l}}&=&\frac{1}{N^2}\sum_{n,m=0}^{N-1}|c_{n,m}(t)|^{2}\nn\\
&&\hspace{-0.2cm}\times\cos\left[\frac{2\pi}{N} \left(n(\mathscr{i}-\mathscr{k})+m(\mathscr{j}-\mathscr{l})\right)\right]\,,
\ea
and
\ba
\Delta(t) &=& ia^2\int_{t_0}^t ds\,m^2(s)\,\delta\phi(s)\nn\\
&&\hspace{0.8cm}\times\left(c_{0,0}(t)^*c_{0,0}(s)-c_{0,0}(t)c_{0,0}(s)^*\right)\,.
\ea

This contains all the information about the quantum dynamics of the field theory we are considering. In the case of domain walls in two or higher dimensions, the relevant quantity to compute is the average area density (strictly speaking length density in two dimensions) of such extended objects. It remains then to write a quantum operator generalizing $\hat{n}_{\rm Z}$ and which, in some limit, will describe the desired observable. A possible choice is
\begin{widetext}
\ba
\hat{\mathcal{A}}_Z&\equiv&\frac{a}{L^2}\sum_{\mathscr{i,\mathscr{j}}=1}^N\frac{1}{4}\left[\rm{sgn}(\hat{\phi}_{\mathscr{i}\mathscr{j}}-\delta\phi(t))-\rm{sgn}(\hat{\phi}_{\mathscr{i}+1,\mathscr{j}}-\delta\phi(t))\right]^2
+\frac{a}{L^2}\sum_{\mathscr{i,\mathscr{j}}=1}^N\frac{1}{4}\left[\rm{sgn}(\hat{\phi}_{\mathscr{i}\mathscr{j}}-\delta\phi(t))-\rm{sgn}(\hat{\phi}_{\mathscr{i},\mathscr{j}+1}-\delta\phi(t))\right]^2\nn\\
&=&\frac{N}{L}-\frac{a}{2L^2}\sum_{\mathscr{i},\mathscr{j}=1}^N \rm{sgn}\left((\hat{\phi}_{\mathscr{i},\mathscr{j}}-\delta\phi)(\hat{\phi}_{\mathscr{i}+1,\mathscr{j}}-\delta\phi)\right)
-\frac{a}{2L^2}\sum_{\mathscr{i},\mathscr{j}=1}^N \rm{sgn}\left((\hat{\phi}_{\mathscr{i},\mathscr{j}}-\delta\phi)(\hat{\phi}_{\mathscr{i},\mathscr{j}+1}-\delta\phi)\right)\,.
\label{operator}
\ea
\end{widetext}

\begin{figure}[t]
      \includegraphics[width=0.45\textwidth,angle=0]{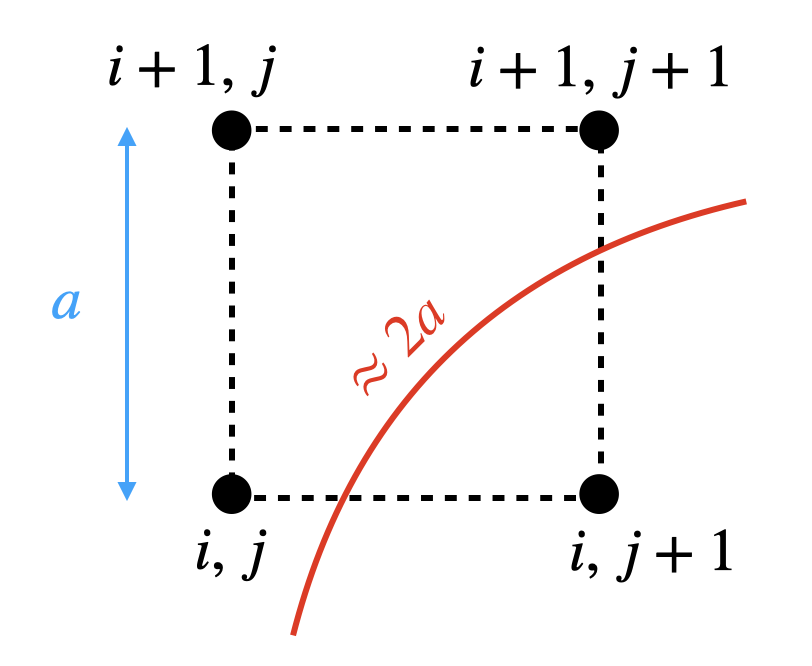}
 \caption{
Sketch of a generic domain wall crossing two adjacent sides of a lattice cell. The field vanishes along the lower and right sides, and the length of the corresponding portion of domain wall is estimated by the operator~\eqref{operator} to be equal to $2a$. Assuming the domain wall to be smooth on scales of the size of the cell, this estimate may differ from the exact value by a factor of at most $\sqrt{2}$.}
\label{plaquette}
\end{figure}

This operator counts the number of sides of cells of our regular square lattice that are traversed by a domain wall {\it i.e.} that are such that the quantity $\phi_{\mathscr{i}\mathscr{j}}-\delta\phi$ changes sign along them, multiplies the result by $a$ (the average length of domain wall traversing a side) and divides it by the total area $L^2$ (see Fig.~ref{plaquette}). Of course it suffers from the same overcounting and undercounting disadvantages as the operator $\hat{n}_{\rm Z}$ but, on top of that, its value can only be trusted up to factors of order 1 since a domain wall that traverses cell diagonally will contribute a length $2a$ instead of $a\sqrt{2}$. The translational and rotational symmetries of the theory (more precisely, the residual symmetries of the discretized theory: discrete translations and rotations by multiples of $\pi/4$) imply that 
\ba
\left\langle\rm{sgn}\left((\hat{\phi}_{\mathscr{i},\mathscr{j}}-\delta\phi)(\hat{\phi}_{\mathscr{i}+1,\mathscr{j}}-\delta\phi)\right)\right\rangle\nn\\
&&\hspace{-1in}
=\left\langle\rm{sgn}\left((\hat{\phi}_{\mathscr{i},\mathscr{j}}-\delta\phi)(\hat{\phi}_{\mathscr{i},\mathscr{j}+1}-\delta\phi)\right)\right\rangle\nn\\
&&\hspace{-1in}
=\left\langle\rm{sgn}\left((\hat{\phi}_{11}-\delta\phi)(\hat{\phi}_{12}-\delta\phi)\right)\right\rangle\,,
\ea
and thus the average value of the operator $\hat{\mathcal{A}}_Z$ simplifies considerably:
\be
\left\langle\hat{\mathcal{A}}_Z\right\rangle=\frac{N}{L}\left[1-\left\langle\rm{sgn}\left((\hat{\phi}_{11}-\delta\phi)(\hat{\phi}_{12}-\delta\phi)\right)\right\rangle\right]\,.
\ee
It is clear from this point onward that the computation will proceed along the same lines as in Sec.~\ref{sec:quantKink}. With the obvious replacements of $\phi_1\rightarrow \phi_{11}$, $\phi_2\rightarrow\phi_{12}$ and the redefinition of
\ba
\alpha(t)&\equiv& [\bm{K}(t)]_{11,11}=\frac{1}{N^2}\sum_{n,m=0}^{N-1}|c_{n,m}(t)|^{2}\,,\\
\beta(t)&\equiv&[\bm{K}(t)]_{11,12}=\frac{1}{N^2}\sum_{n,m=0}^{N-1}|c_{n,m}(t)|^{2}\nn\\
&&\hspace{3.5cm}\times\cos(2\pi m/N)\,,
\ea
we can obtain analogous equations to those in~\eqref{nz2} and~\eqref{nz3}. Introducing the cutoff versions of $\alpha$ and $\beta$ {\it i.e.} restricting the sums to those modes with negative frequency yields $\bar{\alpha}$ and $\bar{\beta}$ and allows us to write the average domain wall area density as 
\ba
\mathcal{A}_{DW} &=& \frac{N}{2L\sqrt{\pi}}\int_0^{\infty}\hspace{-0.3cm}dx\,e^{-x^2}\left[{\rm erf}\left(x\sqrt{\frac{\bar{\alpha}-\bar{\beta}}{\bar{\alpha}+\bar{\beta}}}-\frac{\delta\phi-\Delta}{\sqrt{\bar{\alpha}+\bar{\beta}}}\right)\right.\nn\\
&&\hspace{1.5cm}\left.+{\rm erf}\left(x\sqrt{\frac{\bar{\alpha}-\bar{\beta}}{\bar{\alpha}+\bar{\beta}}}+\frac{\delta\phi-\Delta}{\sqrt{\bar{\alpha}+\bar{\beta}}}\right)\right]\,.
\label{nDW1}
\ea
The no-bias case is analytically integrable and gives
\be
\left.\mathcal{A}_{DW}\right|_{\rm no\, bias}=\frac{N}{\pi L}\tan^{-1}\left(\sqrt{\frac{\bar{\alpha}-\bar{\beta}}{\bar{\alpha}+\bar{\beta}}}\right)
\,,
\label{nDWnobias}
\ee
while, in the $N\rightarrow\infty$ limit, the general result reads
\be
\mathcal{A}_{DW}= \left.\mathcal{A}_{DW}\right|_{\rm no\, bias} \exp\left[-\frac{(\delta\phi-\Delta)^2}{\bar{\alpha}+\bar{\beta}}\right]\,.
\label{nDW2}
\ee
We can also give an analytical estimate of the late time behavior of the average area density of domain walls for the case of a sudden phase transition ($m^2(t)=-m_0^2(2\Theta(t)-1)$ and $\delta\phi(t)=\delta\phi_0\Theta(t)$) in the limit of infinite volume $L\rightarrow\infty$, where
\ba
&&\hspace{-0.4cm}\delta\phi(t)-\Delta(t) = \delta\phi_0\cosh\left(m_0t\right)\Theta(t)\,,\\ 
&&\hspace{-0.4cm}\bar{\alpha}+\bar{\beta} = \frac{1}{2\pi}\bigintsss_{0}^{m_0} \hspace{-0.5cm}kdk\hspace{-0.1cm}\left[\frac{m_0^2\cosh\left[2t\sqrt{m_0^2-k^2}\right]-k^2}{(m_0^2-k^2)\sqrt{m_0^2+k^2}}\right]\,,\\
&&\hspace{-0.4cm}\bar{\alpha}-\bar{\beta} = \frac{a^2}{16\pi}\bigintsss_{0}^{m_0} \hspace{-0.5cm}k^3dk\hspace{-0.1cm}\left[\frac{m_0^2\cosh\left[2t\sqrt{m_0^2-k^2}\right]-k^2}{(m_0^2-k^2)\sqrt{m_0^2+k^2}}\right]\,.
\label{kintmore}
\ea
Indeed, using the above expressions and taking the late time limit we find that
\be
\mathcal{A}_{DW}(t) \sim
\sqrt{\frac{m_0}{t}}\; \exp\left[- 2\pi t\,\delta\phi_0^2 \right]\,.
\ee

In $d+1$ dimensions (where the above results can be extended with minimal modifications) one would get
\be\label{A_DW}
\mathcal{A}_{DW}^{d+1}(t) \sim \sqrt{\frac{m_0}{t}}\; \exp\left[- {\cal C}_d \,\delta\phi_0^2 m_0 \left(\frac{t}{m_0} \right)^{d/2} \right]\,.
\ee
with
\be
{\cal C}_d = 2^{d-1} \,\pi^{d/2}\,.
\ee
Interestingly, for $d=3$ this expression coincides with the result obtained for `phase ordering' in condensed matter \cite{Ohta:1982zz}. We comment on why in Sec.~\ref{discussion}.

\section{VOS model}
\label{sec:VOS}

The  {\em precursor domain walls} introduced above 
are related to the standard DWs created during discrete symmetry breaking transitions, but they differ in an important aspect. Precursor walls are simply the zero iso-surfaces of free tachyonic fluctuating fields and so they don't obey the Nambu Goto (NG) equation. 
The results from Sec.~\ref{sec:dws} imply that these precursor wall network  can also enter a self-similar regime, yet with different properties compared to the standard DW network scaling. Let us try to understand the difference in the language of the effective velocity-dependent one-scale (VOS) models.

First, recall the VOS model for standard DW networks \cite{Avelino:2005kn}. It consists in a simplified description of the DW network in terms of two quantities: $L(t)$, the correlation length or the average separation between  walls; and $v(t)$ the (root-mean-squared) velocity of the walls. Using energy conservation arguments together with general properties of the Nambu-Goto equation of motion \cite{Avelino:2005kn,Kawano:1989mw} for DWs in  $d+1$ dimensions, one arrives at 
\ba\label{VOS}
\dot L &=& \left( 1+d\, v^2 \right) H\, L +  c\, v\,,   \\
\label{VOS2}
\dot v &=& \left( 1- v^2 \right) \left( \frac{k}{L} - d\, H v \right)  \quad {\rm (standard~walls)}
\ea
where $H=\dot a/a$ is the expansion rate and $c$, $k$ are constants -- the so-called energy-loss and momentum parameters respectively. Comparing \eqref{VOS} with \eqref{boltz} (with $n_{\rm K}=1/L$) it is clear that the energy loss parameter $c$ can be interpreted as an effective cross section, for wall-wall interactions. For power-law cosmologies, $a(t)\sim t^\gamma$, these equations lead to an attractor scaling solution where $\dot L$ and $v$ equal a constant, in agreement with field theory numerical simulations of  the networks. 

We can compare this to the equations that control the precursor wall networks. We can obtain these equations by arguing as in Sec.~\ref{sec:kinkDiff}. 
First of all, note that for kinks in $1+1$ dimensions Eqs. \eqref{boltz} and \eqref{v} take the following suggestive form when written in terms of the correlation length ($L=1/n_{\rm K}$ in $1+1$ dimensions),
\ba\label{diffVOS}
\dot L &=& \sigma v\,, \\
\label{diffVOS2}
\dot v &=& -  \kappa \, \frac{ v^2 }{L}\,.
\ea
In $d+1$ dimensions, the correlation length in the DW network is identified as $L\equiv 1 / {\cal A}$ with ${\cal A}$ the physical area density. In terms of this, the equations would be basically unchanged, except that the values of $\sigma, \kappa$ might depend on $d$.

The extension of the results in Sections \ref{sec:kinkDiff} and \ref{sec:dws} to an expanding universe is beyond the scope of this work. 
Yet, in the language of the VOS model \eqref{diffVOS}-\eqref{diffVOS2} the extension seems to suggest itself. 
Neglecting relativistic corrections (as we are interested in diffusive solutions, approaching $v\to0$), the natural expectation is
\ba\label{diffVOSH}
\dot L &=&  H\, L + \sigma v\,,   \\
\label{diffVOSH2}
\dot v &=&  -\kappa \frac{v^2}{L} - d H v   \qquad {\rm (precursor~walls)}
\ea
where $d$ is the number of space dimensions.  

The main difference  between \eqref{diffVOSH}-\eqref{diffVOSH2} and \eqref{VOS}-\eqref{VOS2} is in the tension term, $k/L$, present in \eqref{VOS2}. This tends to increase $v$ in  proportion to the DW curvature, and results from the NG equation of motion \cite{Avelino:2005kn,Kawano:1989mw}. The term is absent in the VOS for the precursor walls, which are subject to some frictional force but no accelerating force from the tension. 

Equations \eqref{diffVOSH}-\eqref{diffVOSH2} have 2 solutions. The trivial one,
\be\label{gas}
L(t) \propto a(t)\,, \qquad v=0\,,
\ee
corresponds to a `gas'  of noninteracting DWs. Wall-wall interactions are  frozen, and  the walls are carried and blown away by the expansion.

The other solution, for a power-law model $a(t)=t^\gamma$, is
\ba\label{curved}
L(t)&\propto& t^\mu\,, \qquad v(t)= v_0 \, t^{\mu-1} \quad {\rm with}\\
\mu&=&\frac{\sigma- (d\, \sigma - \kappa )\,\gamma}{\kappa+\sigma}\,,\\ 
v_0&=&\frac{1-(d+1)\,\gamma}{\kappa+\sigma}~.
\ea
This is the generalization to an expanding universe (and to walls) of the diffusive plasma of kinks of Sec.~\ref{sec:kinkDiff}.
Recall that $v$ is a measure of the magnitude of the typical velocity. The positivity of $v_0$ then implies that (assuming $\kappa+\sigma>0$) this solution exists only for sufficiently slow expansion $\gamma <\gamma_{max}\equiv 1/(d+1)$.
At $\gamma_{max}$, $\mu_{max}=\gamma_{max}$ and so this solution merges with the `DW gas' $L(t) \propto a(t)$ solution. Radiation-domination and matter-domination correspond to $\gamma=2/(d+1)$ and $\gamma=2/d$, too fast an expansion to allow for the nontrivial diffusive scaling.

One must keep in mind, however, that the phenomenological parameters $\kappa$, $\sigma$ can actually depend on the expansion rate, {\em i.e.} on $\gamma$ \cite{Martins:2016ois,Avelino:2019wqd}. As we argue in the next section, there is a reason to expect that the correct scaling for precursor walls in an expanding universe retains the form
$$
L\sim t^{1/2}~,
$$
(that is, $\mu = 1/2$), in terms of the proper cosmic time $t$.  
We can translate this into the following condition for $\kappa$,
\be\label{curved2}
\kappa = \sigma \frac{1- 2 \,d \,p}{1-2\,p}~,
\ee
leading to $v_0=(1-2\,p)/\sigma$. This scaling exists and is distinct from the DW gas only for $p<1/2$. 
This makes sense: the threshold separation between diffusive and DW gas behaviours must be when the expansion $a$ is slower/faster than the diffusion itself. Note that $\kappa$ is negative for $1/2\,d < p < 1/2$. In order to keep up with diffusion $\kappa$ needs to become an accelerating term. 

\section{Discussion}
\label{discussion}

We have presented a new analytic method to compute the annihilation of DW networks driven by {\em population bias} (an asymmetric distribution of the nearly-degenerate vacua). We have used standard QFT methods in flat spacetime to i) identify the appropriate initial condition for the network from the quantum field ground state at the transition, and ii) we have computed the time evolution of the most important quantity that characterizes the DW network, namely, the area density. 



Our main result is the computation of the DW area per unit volume ${\cal A}^{d+1} = ( {\rm DW\, area})/ ({\rm volume})$ in $d+1$ dimensions,
%
%
depicted in Figs.~\ref{resplot0} and \ref{resplot1} for the $d=1$ case. (Qualitatively a similar behaviour occurs in higher dimensions.) Asymptotically, ${\cal A}^{d+1}$ obeys the decay law 
\be\label{R3}
{\cal A}^{d+1} \sim \frac{1}{\sqrt t}\,\exp\left[- (t/t_{ann})^{d/2}\right]~.
\ee
Let us also note that the computation captures an intriguing transient increase which apparently can be understood as nucleation.

Let us comment on a few points. First, note that our flat space result \eqref{ours} agrees with the condensed matter result \cite{Ohta:1982zz}. The method used in \cite{Ohta:1982zz} (which later inspired \cite{Hindmarsh:1996xv}) appears to be radically different from ours but it shares important ingredients. In \cite{Ohta:1982zz} the walls are defined by an auxiliary field $u$ with a statistical averaging with Gaussian statistics, and they move according to a given equation of motion. The equation of motion, however, is not the Nambu Goto (NG) equation but the so-called Allen-Cahn equation \cite{AllenCahn}, which is relevant in friction dominated finite density systems, see \cite{Bray94} for a review. It is a nonlinear dissipative equation which, ignoring nonlinearities, has the structural form $ \dot u = D \,\partial_i^2 u$ with $D$ a constant and $\partial_i^2$ the laplacian.

Our method builds upon a Gaussian relativistic quantum field, not  from an equation of motion for the walls. Instead, we track the {\em precursor walls} (the `zeros' of the field $\phi-\delta\phi(t)$).
%
Such walls are not expected to obey a standard NG equation. Still, at late times the overall motion is expected to  obey non-relativistic scaling, simply because the motion of walls that survive for a long time is encoded in long-wavelength modes of a massive field. Indeed, the non-relativistic limit of a tachyonic massive field is easily obtained by introducing the decomposition $\phi(t,x)= e^{|m| t} \psi(t,x)$, leading to a diffusion-like equation $|m| \dot \psi = \partial_i^2 \psi$  at late times (when  $\ddot \psi \ll |m| \dot \psi$). 

Similary, it is possible to obtain the equation of motion that precursor DWs obey by treating them as semiclassical objects, that is, as DW-like solutions to the classical tachyonic Klein Gordon equation. The analogue to the kink solution is
\be\label{preKink}
\phi = \exp{ \big(\sqrt{k^2+|m^2|} \,t\big)} \sinh{\big(k z\big)}~,
\ee
with arbitrary $k$ and overall amplitude. Indeed, there is a zero in the $z=0$ plane and the exponential time dependence results from the field not being stabilized yet.  It is easy to see that  the ansatz $\exp{ \big(\sqrt{k^2+|m^2|} \,t\big)} \sinh{\big(k [z-z_0(t,x,y)] \big)}$ is also a solution at linear order in the bending $z_0(t,x,y)$ provided it satisfies\footnote{An improved ansatz exists giving an exact bent wall solution that holds everywhere, and which leads to a nonlinear equation for $z_0$. However, it doesn't change much the qualitative properties of \eqref{diffNG} so we do not show it here.}
\be\label{diffNG}
2 \sqrt{k^2+|m^2|} \; \dot z_0 = \partial_i^2\, z_0~.
\ee
Again, the $\ddot z_0$ term has been neglected since we consider late times/long wavelengths.
Thus, the nonrelativistic limit makes these walls obey an Allen-Cahn-like equation, and this explains the agreement with the condensed matter result \cite{Ohta:1982zz} and the origin of the `diffusive' scaling $L\sim t^{1/2}$ of precursor walls.  (Notice that for scaling solutions the value of the diffusion constant factors out.) 

The extension of our results to an expanding FRW cosmology is beyond the scope of this work, however, the reasoning above suggests some expectations. 
Clearly, there are two regimes according to whether the scalar mass is bigger/smaller than the Hubble parameter $H$. For $H\gg |m|$, the field motion is frozen (even for $m^2<0$), and for $H\ll |m|$ we expect precursor walls with non-relativistic limit of the form $\dot z_0 \propto a(t)^{-2}\partial_i^2\, z_0$ with $t$ the cosmic time and $\partial_i$ the comoving spatial gradient. Thus, we expect that without bias there might be a scaling regime where the physical correlation length and velocity keep the same behaviour, $L\sim t^{1/2}$ and $v\sim t^{-1/2}$  in terms of cosmic time. 
This scaling differs from the `DW gas' ($L\sim a(t)$, $v=0$) and it might be realized if the expansion rate is slower than $a(t)\sim t^{1/2}$. The especially interesting case of radiation domination in $3+1$ seems  marginal. In matter domination, $t^{2/3}$, the diffusive scaling $L\sim t^{1/2}$ is not expected to be realized. 
Note also that networks with $L\sim t^{1/2}$ scaling (or close to it) are realized in cosmology if the wall motion is dominated by friction, see \cite{Avelino:2010qf,Martins:2016lzc} and \cite{Blasi:2022ayo} in the context of axionic models.

We can turn now to the network annihilation -- to the exponential decay induced by the bias. Both \eqref{R3} and Hindmarsh's result \eqref{hind} suggest that the exponent is proportional to the correlation volume, $L^d$ (see also \cite{Larsson:1996sp}) -- the comoving volume in the cosmological case \cite{Hindmarsh:1996xv}. This suggests that for a precursor wall network the bias would lead to a suppression of the form
\be\label{ours}
\exp\left[- {\rm const}\, \left( \frac{ \eta   }{  a(\eta) } \right)^{d/2} \right]\,.
\ee
This suppression is much milder than \eqref{hind}, more so for faster expansion rates.
In fact, for $a(t)$ faster than (or equal to) $t^{1/2}$ we would expect DW gas behaviour, $L\sim a(t)$, and the exponent becomes time independent. 

Let us emphasize that \eqref{ours} (if confirmed) is not in contradiction with Hindmarsh's result \eqref{hind}. They simply refer to different regimes: \eqref{hind} holds for stabilized DWs that obey the NG equation whereas \eqref{ours} would apply for precursor walls, that are similar to the DW gas limit.
As mentioned above, a physical precursor wall regime is feasible in cosmology. This should happen for instance if the Hubble rate $H$ at the symmetry breaking transition is significantly larger than the typical mass scale near the symmetry breaking transition. (Another possibility is that the symmetry breaking field is light during inflation.) 
This can result in the formation of an underdense DW network (less than one DW per Hubble patch). Even for stabilized DWs the initial evolution should be close to the DW gas regime $L\sim a(t)$. As in the DW gas limit of precursor walls, a population bias is then also expected not to lead to exponential suppression in time.

It seems, then, that different decay laws, \eqref{coul}, \eqref{hind}, \eqref{ours} (or even different ones), can apply for different network models/realizations.
It is clear from the previous paragraph that a crucial ingredient that specifies the DW network is its initial condition, say, at the symmetry breaking transition.

After all, the network annihilation process consists in the collapse of (fewer and fewer) closed DW structures  larger in size than the horizon at the annihilation time $t_{ann}$. By definition, these are encoded in super-horizon modes, which remain basically frozen during the scaling period. When they evolve, they do so once the scaling period is over. So, it seems rather plausible that the statistics of how many DWs are present after the typical annihilation time is also considerably affected by the initial condition, for deeply super-horizon modes. This picture is indeed confirmed by the recent simulations of \cite{Gonzalez:2022mcx}. Using inflationary initial conditions, the network annihilation is found to be much slower and less sensitive to bias.

It is relevant to compare the assumptions on the initial state also in 
previous works. In Ref. \cite{Hindmarsh:1996xv} a scale invariant (white noise) initial power is assumed. The numerical simulations in \cite{Correia:2014kqa,Correia:2018tty} set the field at the two vacua with (biased) random probability at each lattice sites. It isn't entirely obvious how to map from one to the other. It seems possible that this is why \eqref{hind} fails for the population bias simulations in $2+1$ of \cite{Correia:2014kqa,Correia:2018tty}. In our treatment, the initial condition is encoded in the QFT vacuum of the free massive field. This assumption accounts for the formation of the DWs at the spontaneous symmetry breaking transition.

The present work can be extended in several directions. Whether the guess for the decay \eqref{ours}  really applies in an expanding space, even for precursor walls, requires confirmation. 
It is possible in principle to include different initial states (for instance a thermal state at formation), as well as a different time dependence for both the explicit and spontaneous breaking of the discrete symmetry.
We have considered the simplest model that gives rise to a DW network, with a $\lambda \phi^4$ double-well potential. In principle the method described in this work to count the DW area density can be extended to other models, an interesting target being axionic models. We leave these questions for future work.

\acknowledgements
We thank R. Ferreira, F. Rompineve, T. Vachaspati and M. Mukhopadhyay for useful discussions and comments. 
This work is supported by projects
PID2020-115845GB-I00/AEI/10.13039/501100011033 and 2017-SGR-1069. IFAE is partially funded by the CERCA program of the Generalitat de Catalunya. The work of GZ was supported by a fellowship from “La Caixa” Foundation (ID 100010434) and from the European Union’s Horizon 2020 research and innovation programme under the Marie Skłodowska-Curie grant agreement No 847648. The fellowship code is LCF/BQ/PI20/11760021.

%

\newpage

\bibstyle{aps}
\bibliography{paper}

\end{document}